\documentclass{article}

\usepackage{arxiv}

\usepackage{amsmath,bm}
\usepackage{ulem}
\usepackage{diagbox}
\usepackage{multicol}
\usepackage{amssymb}
\usepackage{graphicx}
\newcount\colveccount
\newcommand*\colvec[1]{
        \global\colveccount#1
        \begin{pmatrix}
        \colvecnext
}
\def\colvecnext#1{
        #1
        \global\advance\colveccount-1
        \ifnum\colveccount>0
                \\
                \expandafter\colvecnext
        \else
                \end{pmatrix}
        \fi
}

\newcommand{\N}{\mathbb{N}} 
\newcommand{\E}{\mathbb{E}}

\title{A Scale Invariant Ranking Function for Learning-to-Rank: A Real-World Use Case}

\author{
  Alessio Petrozziello\thanks{Corresponding Author} \\
  Expedia Group\\
  \texttt{apetrozziello@expediagroup.com} \\
   \And
 Xiaoke Liu\\
 Expedia Group\\
  \texttt{xiaokliu@expediagroup.com} \\
   \And
 Christian Sommeregger\\
 Expedia Group\\
  \texttt{csommeregger@expediagroup.com} \\
}

\begin{document}
\maketitle

\begin{abstract}
Nowadays, Online Travel Agencies provide the main service for booking holidays, business trips, accommodations, etc. As in many e-commerce services where users, items, and preferences are involved, the use of a Recommender System facilitates the navigation of the marketplaces. One of the main challenges when productizing machine learning models (and in this case, Learning-to-Rank models) is the need of, not only consistent pre-processing transformations, but also input features maintaining a similar scale both at training and prediction time. However, the features' scale does not necessarily stay the same in the real-world production environment, which could lead to unexpected ranking order. Normalization techniques such as feature standardization, batch normalization and layer normalization are commonly used to tackle the scaling issue. However, these techniques are sometimes infeasible in real world web applications. First, the normalization of features across thousands of items can significantly impact the website latency. Second, the real-time inference of the item scores is usually distributed across multiple machines on the production environment, thus listwise operations, requiring information from all items to be available, are hard to achieve. To address this issue, in this paper we propose a novel scale-invariant ranking function (dubbed as SIR) which is accomplished by combining a deep and a wide neural network. We incorporate SIR with five state-of-the-art Learning-to-Rank models and compare the performance of the combined models with the classic algorithms on a large data set containing 56 million booked searches from the Hotels.com website. Besides, we simulate four real-world scenarios where the features' scale at the test set is inconsistent with that at the training set. The results reveal that when the features' scale is inconsistent at prediction time, Learning-To-Rank methods incorporating SIR outperform their original counterpart in all scenarios (with performance difference up to 14.7\%), while when the features' scale at the training and test set are consistent our proposal achieves comparable accuracy to the classic algorithms.
\end{abstract}


\section{Introduction}
As in all e-commerce services where users, items, and preferences are involved, a Recommender System (RS) \cite{adomavicius2011context} can be used to facilitate the navigation of big catalogs. Since Online Travel Agencies (OTAs) \cite{ling2014opening, xiang2015adapting, petrozziello2017data, guo2018novel} provide the main service of booking holidays, business trips and accommodations, an adequate recommendation can aid the user in finding the most appealing listing among many. Learning-To-Rank (LTR) techniques can be used to build such RS by training on data consisting of lists of items with some partial order specified among items in each list. LTR aims to learn a scoring function that maps feature vectors to real-valued scores in a supervised setting. Scores computed by such a function induce an ordering of items in the list: the ranking \cite{ai2019learning}. The majority of existing LTR algorithms learn a parameterized function by optimizing a loss that acts on pairs of items (pairwise) \cite{ranknet, lambdarank, lambdarank2} or a list of items (listwise) \cite{listnet, listmle, softrank}. Booking data collected from user logs is an important source of observational data that can be utilized to train LTR algorithms \cite{sorokina2016amazon, karmaker2017application, grbovic2018real, panteli2019recommendation}. However, when used for training, this data not always reflect the same distribution or scale of the inputs received in the e-commerce production environment, leading to unexpected ranking. Many data driven companies employ A/B testing \cite{dixon2011b, cheng2016wide, haldar2019applying} to decide whether a new product or feature is good enough to be rolled out to all customers. In big organizations it is very likely that changes rolled out for testing in different departments would affect the model without the machine learning engineer even knowing about it \cite{sculley2015hidden, iqbal2019production}. Those changes can unwillingly alter the inputs to the LTR model, modifying the sort order of the items on the marketplace. For example, an OTA can change (i.e., A/B test) how the price for a room is displayed to the user: one group of users can be presented with nightly prices, while another group with the price of the full stay. If this scaling issue is not addressed, the two groups of users will see different sort order for the same search query, invalidating the full experiment - i.e., the graphic interface change from nightly price to full price, should not have an impact on the ranking of the listings. Other examples could be: a change in currency, a change of the guest rating range (e.g., from 0-10 to 0-100), or hotel-landmarks distances expressed in kilometers instead of miles. A common solution to this issue is to conduct normalization on the training data, test data and at real-time inference on the production environment. While the first two cases are trivial to handle, real-time normalization is usually unfeasible. The reason is three-fold: Firstly, when a user queries the website, the candidate set usually contains thousands of items (properties listing in our case). Normalization of such large sets take relatively long, which increases the latency of the whole website, negatively impacting the user experience. Secondly, for the sake of latency reduction, the inference of the items score is usually distributed across many machines. Each machine will only be able to access partial information of the list of items, making canonical normalization techniques unreliable. Lastly, listwise standardization doesn’t work if the observed list is truncated, which usually happens in real world production enviroment (for a given query, only a subset of the catalogue is considered for ranking). Since normalization of the input data is hard to achieve, it is necessary to build a LTR algorithm which is invariant to the features scale change.
Here we tackle this problem by proposing a new scale invariant ranking function (dubbed as SIR) that works regardless of the received inputs scale at prediction time. The empirical evaluation demonstrates the usefulness of SIR when incorporated into five state-of-the-art LTR algorithms (namely RankNet, LambdaRank, ListNet, ListMLE and SoftRank) that do not natively address the scale problem at prediction time. We test our solution on a data set containing 56 million booked searches from the Hotels.com website and assess the models' accuracy using the Normalized Discounted Cumulative Gain (NDCG) metric \cite{wang2013theoretical}.
Results show that adding the rank preserving feature does not impact the performance of the models when the features scale of the test set is the same as the one of the training set. However, when a transformation modifying the features' scale is applied to the test set, our proposal is the only one preserving the models' accuracy, while in all but two cases, the original formulations show a significant drop (p-value $<0.001$) in the evaluation metric (up to 14.7\%). This result confirms that the change in scale, if not addressed, is reflected as a change in the ranking order.

\section{Related work}
\subsection{Learning to Rank}
Learning to rank \cite{liu2009learning} is an application of machine learning (usually supervised or semi-supervised) in the construction of ranking models for Information Retrieval systems \cite{manning2008introduction}. LTR methods are classified into three categories: point-wise \cite{dalip2013exploiting, mohan2011web}, pair-wise \cite{ranknet, lambdarank2, lambdarank}, and list-wise \cite{softrank, xu2007adarank, lan2014position, listnet}.

Point-wise approaches only look at a single item in the loss function. They consider each item individually and train a classifier/regressor on it to predict how relevant it is for the current query (i.e., search). The final ranking is achieved by simply sorting the result list by these items score. For point-wise approaches, the score for each item is independent of the other items that are in the results list for the query. All the standard regression and classification algorithms \cite{dalip2013exploiting, mohan2011web} can be directly used for point-wise learning to rank.

Pair-wise approaches look at a pair of items at a time in the loss function. Given a pair of items, they try and come up with the optimal ordering for that pair and compare it to the ground truth. The goal for the ranker is to minimize the number of inversions in ranking (i.e. cases where the pair of results are in the wrong order relative to the ground truth).
Pair-wise approaches work better in practice than point-wise approaches because predicting relative order is closer to the nature of ranking than predicting class label or relevance score \cite{listnet}. Some of the most popular LTR algorithms like RankNet \cite{ranknet}, LambdaRank \cite{lambdarank2} and LambdaMART \cite{lambdarank} are pair-wise approaches.

List-wise approaches directly look at the entire list of items and try to come up with the optimal ordering for it. There are two main sub-techniques for doing list-wise LTR: direct optimization of Information Retrieval measures such as NDCG (e.g. SoftRank \cite{softrank}, AdaRank \cite{xu2007adarank}); minimization of a loss function that is defined based on understanding the unique properties of the kind of ranking of the task at hand (e.g. ListNet \cite{listnet}, ListMLE \cite{lan2014position}).
List-wise approaches can get fairly complex compared to point-wise or pair-wise approaches and are usually yielding better ranking performance due to the fact that the features of the whole list are considered at training time. In this paper we incorporate two pairwise learning-to-rank methods (Ranknet and LambdaRank) and three listwise algorithms (Listnet, ListMLE and Softrank) with the proposed scale invariant ranking function.

\subsection{Related Work}
It is widely acknowledged that Recommender Systems should be invariant with repsective to the scale of object features and the scale of customer ratings \cite{pennock2000social}. Unfortunately, the classic RS models usually depend on such scale. For example, the computation of user similarity in collaborative filtering (CF) highly depends on the scale of customer ratings, while different users usually use different rating scales \cite{yan2013exploiting}.

A great research effort has been done in the last two decades to tackle the features scale problem for CF. To avoid the mismatch in the users' internal ratings scale, Yu and Yang \cite{yu2008collaborative} proposed to use ordinal information rather than absolute ratings when computing user similarity. Lemire \cite{lemire2005scale} introduced an adjusted pearson correlation similarity, Su and Khoshgoftaar \cite{su2009survey} presented an adjusted cosine similarity. Both similarity metrics are invariant to the scale of customer ratings.  Anand and Bharadwaj \cite{anand2010adaptive} suggested to transform the raw customer ratings to an optimal rating via genetic programming and then use the transformed ratings to determine user similarity.

Besides, researchers employ normalization techniques to adapt to various rating habits of users, normalising ratings assigned by different users to the same scale by some criteria \cite{xie2015mathematical}. For example, Resnick et.al \cite{resnick1994grouplens} developed a $z$-score normalization based method, Sarwar et.al \cite{sarwar2000application} proposed a dimensionality reduction based method, Lemire \cite{lemire2005scale} introduced an $L_p$ normalization method, and Traupman and Wilensky \cite{traupman2004collaborative} presented a factor analysis based method.

Nowadays, Deep Neural Network (DNN) based models are widely used in Recommender Systems. The scale issue of DNN models has been identified through the literature. For example, DNN with ReLU activation function was proved to be scale invariant regarding to the weight space. If the ingoing weights of one hidden node are multiplied by a constant $c$ and the outgoing weights of the hidden node are divided by $c$ at the same time, the output of network is identical \cite{yi2019positively, neyshabur2015path}. The weight space invariance implies that there can be an infinite amount of parameter points in the weight space that represent the same model \cite{yuan2019scaling}.

The scale invariant property of DNN models regarding to the input features are mostly done by normalization. The scale of features can be unified by canonical normalization. For example, the scale of numerical features can be easily unified by transforming them into normal distributed variables with mean 0 and variance 1. Since Salimans and Kingma \cite{salimans2016weight} proposed the batch normalization, researchers proposed a variety of normalization techniques to speed up the training of neural networks and tackle the scaling issue, including weight nomalization \cite{salimans2016weight}, layer normalization \cite{ba2016layer} and streaming normalization \cite{liao2016streaming}. In batch normalization, the pre-activation of each neuron is normalized by the mean and the standard deviation of the outputs computed over the samples in the mini batch \cite{yuan2019scaling}. Adding batch normalization to the feed-forward neural networks can improve training speed. Layer normalization is proposed to overcome the disadvantage of batch normalization, where pre-activation of each neuron is normalized by all the hidden neurons in the same layers instead of the pre-activation of each neuron within a mini batch. While those proposed techniques are used to speed up the training of models, they are not feasible at inference time for several reasons. Firstly, the normalization of large sets (i.e., online queries can retrieve up to thousands of items at each time) negatively impacts the environment latency . Furthermore, items in the same search are usually handled and scored by different machines, meaning each machine only has partial information, making the normalization hard to achieve.

\section{Proposed method}
\label{proposed}
In this paper we propose a neural network based ranking function which takes a list of items (represented by some item features and query features) in a query as input and returns a sorted list as output. Before diving into the formal explanation of the proposed method we clarify the notations we use throughout the paper.\par
\footnotesize \makebox[2cm]{$\bm x_{ij}\in\mathbb{R}^{M+K_1+K_2}$}\normalsize Feature vector of the item $j$ in query $i$. It can be split into query features $\bm x_{i}^Q$ and item features $\bm x_{ij}^{I}.$\par
\small \makebox[2cm]{$\bm x_{i}^Q\in\mathbb{R}^M$}\normalsize The query feature vector of query $i$. Items in the same query share same query features.\par
\small \makebox[2cm]{$\bm x_{ij}^{I} \in \mathbb{R}^{K_1+K_2}$}\normalsize Item feature vector of item $j$ in query $i$. It can be further split into $\bm x_{ij}^{I,F}$ and $\bm x_{ij}^{I,S}$.\par
\small \makebox[2cm]{$\bm x_{ij}^{I,F} \in \mathbb{R}^{K_1}$}\normalsize The partition of item features whose scale remain the same in all cases.\par
\small \makebox[2cm]{$\bm x_{ij}^{I,S} \in \mathbb{R}^{K_2}_{++}$}\normalsize The partition of item features whose scale can vary at prediction time (to which our method is scale invariant). Note that the partition of features which have scaling issue is known. Here $\mathbb{R}_{++}$ denotes positive real numbers.\par
\small \makebox[2cm]{$y_{ij}$}\normalsize Customer label of item $j$ in query $i$ where $y_{ij}$ = 1 for the booked item and $y_{ij}$ = 0 for others.\par
\small \makebox[2cm]{$D_i$}\normalsize The number of items in query $i$.\par
\small \makebox[2cm]{$N$}\normalsize The total number of queries.\par
\normalsize

\begin{figure*}[h]
\centering
\includegraphics[width=0.9\textwidth]{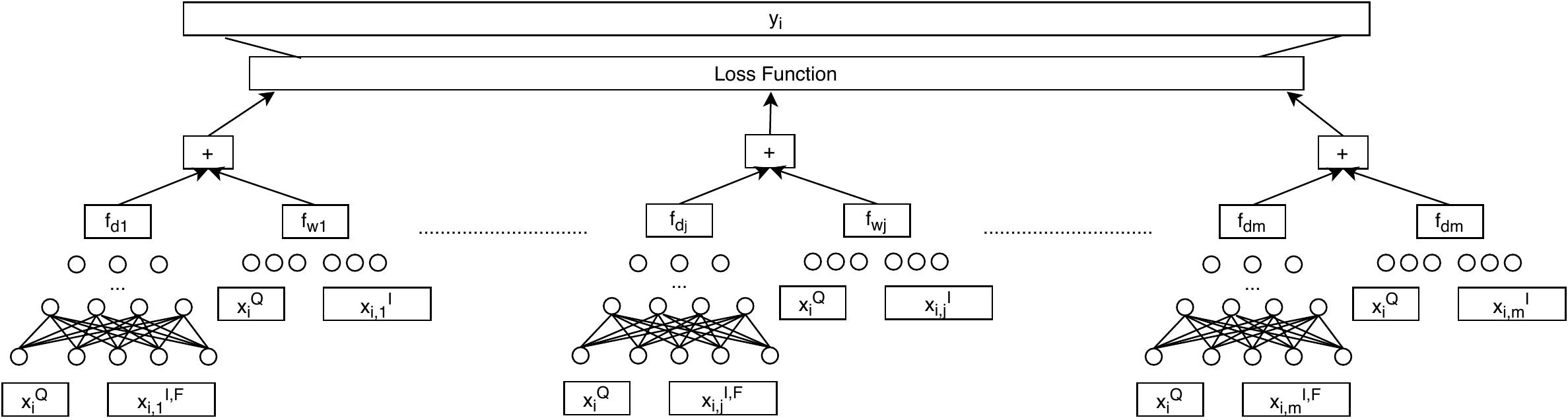}
\caption{The SIR architecture is a siamese network with $m$ (i.e., number of items in each list) identical branches (i.e., shared weights), each composed of two parts: a deep part taking as input the query features ($x^{Q}_{i}$) and the scale invariant item features ($\bm x^{I,F}_{ij}$), and a wide part taking the query features ($x^{Q}_{i}$) and both scale variant and invariant item features ($\bm x^{I}_{ij}$). Each part produces a score: $f_d(\bm x^Q_{i},\bm x_{ij}^{I,F})$ and $f_w(\bm x^Q_{i},\bm x_{ij}^{I})$ for the deep and wide part respectively (denoted as $f_d$ and $f_w$ in the figure), which are then added together.
When testing the original LTR methods, only the deep part of the structure is used, taking all item features as input ($\bm x^{I}_{ij}$). The query features ($x^{Q}_{i}$) by nature do not suffer of the scale variance problem.\vspace{-15pt}}
\label{fig:approach}
\end{figure*}

The proposed scoring function $f_{n}$ is composed of two parts, a deep part $f_{d}$ which takes $\bm x_{i}^{Q}$ and $\bm x^{I,F}_{ij}$ as input and calculates the score of the items using a Deep Neural Network with three fully-connected layers (512, 256, and 128 neurons respectively) and a dense layer of 25 elements containing the scores (see more details in Figure \ref{fig:approach}); and a wide part $f_{w}$ which takes $\bm x_{i}^{Q}$ and $\bm x^{I}_{ij}$ as input, and linearly interacts them through their outer product. The produced score is then added to the score produced by $f_{d}$, and passed to a soft-max function to convert the scores into probabilities. Details on the network structure can be found in Figure \ref{fig:approach}. Specifically, the score of item $j$ in query $i$:
\begin{align}\label{scoringfunction}
&f_{n}(\bm x_{ij}) = f_d(\bm x^Q_{i},\bm x_{ij}^{I,F}) + f_w(\bm x^Q_{i},\bm x_{ij}^{I}),\\
&\text{where }f_w(\bm x^Q_{i},\bm x_{ij}^{I}) = <\bm w,
(\bm f_s(\bm x^{Q}_{i}) \otimes_{kron} \log(\bm x^{I}_{ij}))>,\notag\\
&f_{n}: \mathbb{R}^{M+K_1+K_2} \rightarrow \mathbb{R}, \;\;\;\;\;
f_d  :\mathbb{R}^M \times \mathbb{R}^{K_1} \rightarrow \mathbb{R},\notag\\
&f_w:\mathbb{R}^M \times \mathbb{R}^{K_1+K_2} \rightarrow \mathbb{R},\;\;\;\;\;\bm f_s  :\mathbb{R}^{M} \rightarrow \mathbb{R}^L, \;\;\;\;\;
\bm w \in \mathbb{R}^{KL}.\notag
\end{align}
Here $\bm f_{s}$ compresses the query features by projecting them to $\mathbb{R}^L$ with $L < M$, $\otimes_{kron}$ is the kronecker product, $\bm w$ is a weight parameter to tune, and $<\cdot>$ represents the inner product operator. 

Then for two items $\bm x_{ij}$ and $\bm x_{ik}$ in the same query, the score difference between $\bm x_{ij}$ and $\bm x_{ik}$:

\begin{align}\label{scaleinvariant0}
f_n(\bm x_{ij}) - f_n(\bm x_{ik}) &= \left (f_d(\bm x^Q_{i},\bm x_{ij}^{I,F}) + f_w(\bm x^Q_{i},\bm x_{ij}^{I})\right )\notag\\
&- \left (f_d(\bm x^Q_{i},\bm x_{ik}^{I,F}) + f_w(\bm x^Q_{i},\bm x_{ik}^{I})\right )\notag\\
&=\left (f_d(\bm x^Q_{i},\bm x_{ij}^{I,F}) + f_w(\bm x^Q_{i},[x_{ij}^{I,F},x_{ij}^{I,S}]\right )\notag\\
&- \left (f_d(\bm x^Q_{i},\bm x_{ik}^{I,F}) + f_w(\bm x^Q_{i},\bm [x_{ik}^{I,F},x_{ik}^{I,S}])\right ).
\end{align}
\normalsize
Now assume the scale of \small{$\bm x_{i}^{I,S}$}\normalsize changes to $c$ times ($c>0$), i.e. \small{$\tilde{\bm x}_{ij}^{I,S} = c\bm x_{ij}^{I,S}$}\normalsize and \small{$\tilde{\bm x}_{ik}^{I,S} = c\bm x_{ik}^{I,S}$}\normalsize where $c$ is a scaler which can take any positive value, then the score difference of these two items after the scale change:
\begin{align}\label{scaleinvariant1}
f_n(\tilde{\bm x}_{ij}) - f_n(\tilde{\bm x}_{ik})
&=\left (f_d(\bm x^Q_{i},\bm x_{ij}^{I,F}) + f_w(\bm x^Q_{i},\tilde{\bm x}_{ij}^{I})\right )\notag\\
&- \left (f_d(\bm x^Q_{i},\bm x_{ik}^{I,F}) + f_w(\bm x^Q_{i},\tilde{\bm x}_{ik}^{I})\right ).\notag\\
&=\left (f_d(\bm x^Q_{i},\bm x_{ij}^{I,F}) + f_w(\bm x^Q_{i},[x_{ij}^{I,F},cx_{ij}^{I,S}]\right )\notag\\
&- \left (f_d(\bm x^Q_{i},\bm x_{ik}^{I,F}) + f_w(\bm x^Q_{i},[x_{ik}^{I,F},cx_{ik}^{I,S}])\right ).
\end{align}
\normalsize
Now we show that the score difference between $\bm x_{ij}$ and $\bm x_{ik}$ does not change before and after scaling, i.e. $f_n(\tilde{\bm x}_{ij}) - f_n(\tilde{\bm x}_{ik})=f_n(\bm x_{ij}) - f_n(\bm x_{ik})$.

If we denote the concatenation of vectors using $\oplus$, formula (\ref{scaleinvariant1}) can be expanded as follows,
\begin{align}\label{scaleinvariant2}
&f_n(\tilde{\bm x}_{ij}) - f_n(\tilde{\bm x}_{ik})\notag\\
=&\left (f_d(\bm x^Q_{i},\bm x_{ij}^{I,F}) + f_w(\bm x^Q_{i},\tilde{\bm x}_{ij}^{I})\right ) - \left (f_d(\bm x^Q_{i},\bm x_{ik}^{I,F}) + f_w(\bm x^Q_{i},\tilde{\bm x}_{ik}^{I})\right ),\notag\\
=&f_d(\bm x^Q_{i},\bm x_{ij}^{I,F}) + <\bm w,
\left (\bm f_s(\bm x^{Q}_{i}) \otimes_{kron} \log(\tilde{\bm x}^{I}_{ij})\right )>\notag\\
&- f_d(\bm x^Q_{i},\bm x_{ik}^{I,F}) - <\bm w,\left (\bm f_s(\bm x^{Q}_{i}) \otimes_{kron} \log(\tilde{\bm x}^{I}_{ik})\right )>,\notag\\
=&f_d(\bm x^Q_{i},\bm x_{ij}^{I,F}) + <\bm w,
\left (\bm f_s(\bm x^{Q}_{i}) \otimes_{kron} (\log(\bm x^{I,F}_{ij})\oplus\log(c\bm x^{I,S}_{ij}))\right )>\notag\\
&- f_d(\bm x^Q_{i},\bm x_{ik}^{I,F}) - <\bm w,\left (\bm f_s(\bm x^{Q}_{i}) \otimes_{kron} (\log(\bm x^{I,F}_{ik})\oplus\log(c\bm x^{I,S}_{ik}))\right )>,\notag\\
=&f_d(\bm x^Q_{i},\bm x_{ij}^{I,F}) + <\bm w,
\Big (\bm f_s(\bm x^{Q}_{i}) \otimes_{kron} (\log(\bm x^{I,F}_{ij})\oplus\log(\bm x^{I,S}_{ij}))\Big )>\notag\\ 
&+ <\bm w,\Big (\bm f_s(\bm x^{Q}_{i}) \otimes_{kron} (\log(\bm x^{I,F}_{ij})\oplus\log(c))\Big )>-f_d(\bm x^Q_{i},\bm x_{ik}^{I,F}) \notag\\
&- <\bm w,\Big (\bm f_s(\bm x^{Q}_{i}) \otimes_{kron} (\log(\bm x^{I,F}_{ik})\oplus\log(\bm x^{I,S}_{ik}))\Big )>\notag\\
&- <\bm w,
\Big (\bm f_s(\bm x^{Q}_{i}) \otimes_{kron} (\log(\bm x^{I,F}_{ij})\oplus\log(c))\Big )>,\notag\\
=&f_d(\bm x^Q_{i},\bm x_{ij}^{I,F}) + <\bm w,
(\bm f_s(\bm x^{Q}_{i}) \otimes_{kron} \log(\bm x^{I}_{ij}))>\notag\\ 
&- f_d(\bm x^Q_{i},\bm x_{ik}^{I,F})- <\bm w,
(\bm f_s(\bm x^{Q}_{i}) \otimes_{kron} \log(\bm x^{I}_{ik}))>\,\notag\\
=&f_n(\bm x_{ij}) - f_n(\bm x_{ik}).
\end{align}
In short,
\begin{equation}\label{scaleinvariant2}
f_n(\tilde{\bm x}_{ij}) - f_n(\tilde{\bm x}_{ik})=f_n(\bm x_{ij}) - f_n(\bm x_{ik}).   
\end{equation}
Namely, the proposed scoring function gives the same score difference for two items in the same search with or without scaling. The difference of scores is invariant to the scale change. Accordingly, the ranking based on this scoring function is scale invariant as well.


In query $i$, assume we have scores ($f(\bm x_{i1})$, $f(\bm x_{i2})$, $\dots$, $f(\bm x_{iD_i})$), then a ranking function is a bijective function from  $\{1, 2, \dots, D_i\}$ to $\{1, 2, \dots , D_i\}$, namely
\begin{equation}
\pi_i:\{1, 2, \dots, D_i\} \rightarrowtail\!\!\!\!\!\rightarrow \{1, 2, \dots , D_i\},
\end{equation}
where $\rightarrowtail\!\!\!\!\!\rightarrow$ indicates bijection. Here we rank items according to scores in decending order, i.e. our ranking function $\pi$ satisfies $f(\bm x_{i\pi_i^{-1}(1)}) >= f(\bm x_{i \pi_i^{-1}(2)})>=\ldots>=f(\bm x_{i\pi_i^{-1}(D_i)})$. Here by $\pi_i^{-1}(j)$ we mean the index of the item ranked in the $j$-th place in the query by the ranking function $\pi_i$.

The proposed method has some similarities to the NN\&FM ensemble introduced in \cite{haldar2019applying} and Wide\&Deep approach presented in \cite{cheng2016wide}, however the main goal here is not to improve the model accuracy but to restrict the structure of our model to allow for more robust behavior when used in a production environment.

\section{Empirical Study Design}

This section presents the design of the empirical study we carried out to get an insight into the use of a ranking preserving function for Learning-To-Rank tasks. We first present the research questions we aim to answer and then the data and techniques we experimented with, and the evaluation criteria we used to assess the results.

\subsection{Research Questions}
Before checking whether the proposed approach is useful in solving the scale variance problem, we must ensure it yields comparable results to the current state-of-the-art techniques when there is no scale variance in the test set. Thus, the first research question we aim to answer is:
RQ1. Does our proposal provide at least comparable ranking accuracy than currently used LTR methods when training and test set features' scale is consistent?
To answer RQ1 we incorporate our proposal into five widely-used and well-studied state-of-the-art LTR methods, namely RankNet \cite{ranknet}, LambdaRank \cite{lambdarank}, ListNet \cite{listnet}, ListMLE \cite{listmle} and SoftRank \cite{softrank}, as detailed in Section \ref{benchmarks}. If SIR models achieve at least comparable performance to the original LTR algorithms, then it could be employed in a production environment without losing any ranking accuracy.
Following RQ1, we want to know whether the proposal outperforms the state-of-the-art techniques when the scale variance problem in the test set is considered. This motivates our next research question:
RQ2. Is the proposal invariant to test set features' scale changes?
To answer RQ2 we perturb the scale of some test set features based on real-world scenarios, and compare again ranking accuracy of models incorporating SIR, to their scale variant counterpart. To validate our hypothesis we simulate four cases which could happen in a real-world e-commerce production environment. \textit{Case 1}: The hotel price and discount passed to the model at prediction time are the full price/discount of the stay (price/discount of 1 night, times the number of nights), instead of the nightly price/discount. \textit{Case 2}: The hotel price and discount respectively are passed to the model at prediction time in the local currency, instead of dollars (USD). To achieve that, we multiply the nightly price and discount by the daily exchange rate of the selected currency. \textit{Case 3}: We combine \textit{Case 1} and \textit{Case 2}, hence the price and discount passed to the model at prediction time are the full price/discount of the stay, in the local currency. \textit{Case 4}: We select a currency with a very high exchange rate compared to the dollar (i.e., Korean Won, 1 USD = 1200 KWR) and simulate what would happen if all test set searches were performed from the Korean point of sale (this case is less realistic than the first three, and only used to check performance of the models in case of a big shift in the scale of the features - i.e., price and discount).

\subsection{Data set}

To carry out the empirical study we exploited a travel accommodation booking data set from the Hotels.com website. The data set has 56 million of booked searches, each of which contains up to 25 items (i.e., properties). Each item has 16 features shared (i.e., query features) with all the others in the same search (e.g., destination, check in date, check out date, point of sale, locale, local currency, user device, etc.), and 14 item features (e.g., room price, discount, star rating, review scores, point of supply, etc.). Each item also has one boolean target variable representing whether it has been booked. The task is to rank the items such that the booked one has the largest score (i.e., is ranked higher) based on the given query and item features. 
Before feeding the data to the ranking algorithms, we apply widely used features transformations to facilitate the learning procedure: All numerical features are standardized ($(feature\_val - \mu) / \sigma$), where $\mu$ is the feature mean and $\sigma$ the standard deviation; while all categorical features are indexed (i.e., 1 to \#levels) and mapped to an embedding learned within the Neural Network at training time. This technique allows to deal with high cardinality features in a simple fashion \cite{haldar2019applying}. 

\subsection{Validation and Evaluation}
\label{validation}
We use a data set containing 56 million of booked searches from Hotels.com website to verify our proposal. In each search only one property is labeled as the booked one according to the customer's booking record, while the others are labeled as non-booked. To verify the efficiency of the proposed method a validation process is required. We used a hold-out technique \cite{raschka2018model}, where the data is split in two disjoint sets, the first one used for training the model, the second one for evaluating its performance. We randomly split the data into a training set with 70\% of data (38.68 million searches) and a test set with 30\% data (16.90 million searches). To avoid over fitting during the training procedure, we use 10\% of the training data (3.86 mission searches) as a separate validation set, and stop the training procedure when either max training iteration is reached (i.e., 100) or the validation NDCG does not improve for 20 consecutive iterations. 

To guarantee consistency across the experiments we use the same proposed network structure (Figure \ref{fig:approach} and scoring function (Eq.~\ref{scoringfunction}). Furthermore, all trained models are equally tuned with classic hyper-parameter optimization \cite{bergstra2013hyperopt}.
 
Concerning the evaluation of our proposed method, we employ one of the most widely used measure in Learning-To-Rank, the NDCG \cite{wang2013theoretical}. The NDCG is a normalization of the Discounted Cumulative Gain (DCG) while DCG is a weighted sum of the degree of relevancy of the ranked items.

Specifically, for given query $i$ and associated $D_i$ items, suppose that $\pi_i$ is a ranking (permutation) on $D_i$ items. Then DCG at position $l$ is defined as:

\begin{equation}
DCG_i(l) = \sum_{j:\pi_i(j)\leqslant l} G_i(j)\,D(\pi_i(j)),
\end{equation}
where
\begin{equation}
G_i(j) = 2^{y_{ij}}-1,\;\;\;\;\; D(\pi_i(j)) = \frac{1}{\log_2(1 + \pi_i(j))}.
\end{equation}
Here $G(\cdot)$ is a gain function, $D(\cdot)$ is a position discount function, $\pi_i(j)$ is the position of item $j$ in ranking list $\pi_i$.

Namely, DCG at position $l$ is:
\begin{equation}
DCG_i(l) = \sum_{j:\pi_i(j)\leqslant l} \frac{2^{y_{ij}}-1}{\log_2(1 + \pi_i(j))},
\end{equation}
while NDCG is a normalization of DCG. NDCG at position $l$ is defined as:
\begin{equation}
NDCG_i(l) = G^{-1}_{\max, i}(l)\sum_{j:\pi_i(j)\leqslant l} \frac{2^{y_{ij}}-1}{\log_2(1 + \pi_i(j))}.
\end{equation}
Here $G_{\max, i}(l)$ is the normalizing factor and is chosen such that a perfect ranking $\pi_i^{*}$’s NDCG score at position $l$ is 1. By a perfect ranking we mean in this ranking items with higher grades are always ranked higher.

In this paper we set $l=D_i$ and use NDCG of the full list to evaluate ranking performance. We take the average NDCG across all queries on the test set as a measurement of the model performance. We compare the average NDCG of our models and five state-of-the-art models using two sample $t$-test. To be specific, the null hypothesis to be tested is: the NDCG provided by model $M (SIR) $ is significantly smaller than the NDCG provided by model $M$, (when $M (SIR)$ and $M$ are the scale invariant version and original formulation respectively), under significance level $\alpha$=0.05, applying Bonferroni correction for multiple hypothesis testing ($\alpha$/n, where n (n = 25) is total number of comparisons).

\subsection{Benchmarks}
\label{benchmarks}
Learning-To-Rank methods can be classified into three categories: point-wise \cite{dalip2013exploiting, mohan2011web}, pair-wise \cite{ranknet, lambdarank2, lambdarank}, and list-wise \cite{softrank, xu2007adarank, listnet, lan2014position}. In this work, we compare our proposed methods with five widely used LTR technique, which do not natively address the scale invariance problem, two from the pair-wise family (namely RankNet and LambdaRank) and three from the list-wise family (namely ListNet, ListMLE, and SoftRank). Point-wise methods have not been considered as they in general lead to inferior results than pair-wise and list-wise methods \cite{liu2011learning}.

\subsubsection{RankNet}
RankNet is a Neural Network based learning to rank algorithm with a pair-wise loss proposed by Burges et.al. in 2005 \cite{ranknet}. It trains a back-propagation deep neural network with a pairwise loss. 

Specifically, for a pair of items denoted by $\bm x_{ij}$ and $\bm x_{ik}$ in query $i$, we denote the relationship that item $\bm x_{ij}$ is ranked higher than $\bm x_{ik}$ as $\bm x_j \triangleleft \bm x_k$, and the relationship $\bm x_{ij}$ and $\bm x_{ik}$ are in the same relevance class as $\bm x_{ij} == \bm x_{ik}$. The modeled posterior probability that sample $\bm x_{ij}$ is ranked higher than sample $\bm x_{ik}$ is denoted as ${P}_{i,jk}$. Let $\bar{P}_{i,jk}$ be the desired target values for this posterior probability. Then the loss for the pair $\bm x_{ij}$ and $\bm x_{ik}$ is defined by a cross entropy function:
\begin{align}\label{ranknetloss}
l_{i,jk} = -\bar{P}_{i,jk}\log P_{i,jk} - (1-\bar{P}_{i,jk})\log(1-P_{i,jk}),
\end{align}
where
\begin{align}
\bar{P}_{i,jk}=
\left\{
\begin{matrix}
1, \text{ if } \bm x_{ij}\triangleleft \bm x_{ik}\\ 
0.5,  \text{ if } \bm x_{ij} == \bm x_{ik}\\
0, \text{ if } \bm x_{ik}\triangleleft \bm x_{ij}
\end{matrix}\right., \;\;\;\;\;P_{i,jk}  = \frac{e^{d_{jk}}}{1+e^{d_{jk}}},\;\;\;\;\; d_{jk} = f(\bm x_{ij}) - f(\bm x_{ik}).  
\end{align}
Here $f(\bm x_{ij})$ and $f(\bm x_{ik})$ are the scores of item $\bm x_{ij}$ and item $\bm x_{ik}$ from the network.

Then the total RankNet loss is defined as:
\begin{align}
\bm L = -\sum_{i=1}^{N} \sum_{j \neq k}\left [ \bar{P}_{i,jk}\log P_{i,jk} + (1-\bar{P}_{i,jk})\log(1-P_{i,jk})\right],
\end{align}
for query $1 \leqslant i \leqslant N$ and all pairs of item $\bm x_{ij}$ and $\bm x_{ik}$ in query $i$. 

In the implementation of RankNet, due to the lack of supervised information for items in the same relevance class, we do not consider loss between tied items (loss between non-booked items). In other words we only consider loss between the booked one and the non-booked ones in each search.

\subsubsection{LambdaRank}
LambdaRank is an enhancement of the RankNet algorithm proposed by Chris et. al. in 2010 \cite{lambdarank}, which incorporates the evaluation metric in the learning procedure. The basic idea is to dynamically adjust the loss during the training based on the ranking evaluation metrics. Using NDCG as an example, $\Delta$NDCG is defined as the absolute difference between the NDCG values when two items $j$ and $k$ are swapped. LambdaRank updates the RankNet loss in equation (\ref{ranknetloss}) by reweighing the loss of each item pair by $\Delta$NDCG in each iteration. Namely, the lambdarank loss for item $j$ and item $k$ in query $i$:
\begin{equation}
l_{i,jk} = |\Delta NDCG(j,k)| l_{i,jk}^{\text{ranknet}}, \end{equation}
where
\begin{equation}
|\Delta NDCG(j,k)| = |\bm{G}_{jk} - \bm{G}_{kj}|\cdot|\frac{1}{\bm{D}_{jk}} - \frac{1}{\bm{D}_{kj}}|.
\end{equation}
Here $l_{i,jk}^{\text{ranknet}}$ is the ranknet loss for the corresponding item pair. Then the total lambdarank loss:
\begin{equation}
L = -\sum_{i=1}^{N} \sum_{j \neq k} l_{i,jk}.
\end{equation}
Neural network model trained using the above loss is called lambdarank \cite{lambdarank2}.

\subsubsection{ListNet}
ListNet is a Neural Network based learning to rank algorithm with a list-wise loss, proposed by Zhe Cao et.al. in 2007 \cite{listnet}. ListNet trains a back-propagation deep Neural Network with categorical cross-entropy loss:
\begin{align}
L = -\sum_{i=1}^{N}\sum_{j=1}^{D_i}\bar{P}_{i,j}\log P_{i,j},
\end{align}
where the target probability $\bar{P}_{i,jk}$ = $\frac{e^{y_{ij}}}{\sum_{j=1}^{j=D_i}e^{y_{ij}}}$, the posterior probability from the model $P_{i,jk}$ = $\frac{e^{f(\bm x_{ij})}}{\sum_{j=1}^{j=D_i}e^{f(\bm x_{ij})}}$, where $f(\bm x_{ij})$ is the score of item $\bm x_{ij}$.

\subsubsection{ListMLE}
ListMLE, proposed by Xia et.al. \cite{listmle}, is a Neural Network based LTR algorithm with a listwise negative likelihood loss. The likelihood loss is defined as:
\begin{equation}
L = -\sum_{i=1}^{N}\log P(\bm{y}_{i}|\bm x_{i}, f),
\end{equation}
where
\begin{equation}
P(\bm{y}_{i}|\bm x_{i}, f) = \sum_{j=1}^{D_{i}}\frac{\exp({ f(\bm x_{y_i(j)})})}{\sum^{D_i}_{j}\exp({{f(\bm x_{y_i(k)})}})}.
\end{equation}
Here $\bm x_{i}$ represents the items in query $i$, $f$ is the scoring function, $\bm y_i$ represents the user-assigned relevance ranking of query $i$. $y_i(j)$ is the index of the object which is ranked at position $j$ in query $i$. 

Same as RankNet, due to the lack of supervised information between items in the same relevance level, in the implementation of ListMLE we do not consider loss between tied items.

\subsubsection{SoftRank}
SoftRank is a Neural Network based learning to rank algorithm which was proposed by Taylor et. al. in 2008 \cite{softrank}. Instead of training on a proxy loss function, SoftRank directly optimizes the NDCG metric. However, since NDCG is non-smooth with respect to the ranking scores, a smoothed approximation to NDCG was proposed and then used to obtain the best rank. The main idea of SoftRank is to replace the discount function in NDCG that is discontinuous with respect to the scores with a function continuous to the scores and directly optimize this adjusted NDCG. 
In the SoftRank loss function there is one additional hyper-parameter ($\sigma$) indicating the standard deviation of the score distribution, which was tuned by cross-validation and the best value ($\sigma = 0.15$) was employed. Furthermore, for computation simplicity, we train the SoftRank on lists of 9 items (8 non-booked items and the booked one) instead of the original list of 25. The training time increases exponentially with the growing size of the lists 
while the accuracy does not significantly increases when using more than 9 items. The 8 out of 24 negative samples are chosen randomly. 

Specifically, assume the true score of the item $j$ in query $i$, $s_{ij}$ $\sim$ $\N$ ($z_{ij}$, $\sigma^2$), where $z_{ij}$ is the predicted score, namely
\begin{align}
    P(s_{ij} = s) = \frac{1}{\sqrt{2\pi}\sigma}\exp(-\frac{(s-z_{ij})^2}{2\sigma^2}).
\end{align}
Here $\sigma$ is a hyperparameter to tune indicating the variability of the score. Then for any pair of items $j$ and $k$ in query $i$, the true score difference between them follows a normal distribution as well, and the probability that item $j$ ranks higher than item $k$:
\begin{equation}
d_{jk} = (s_{ij}-s_{ik}) \sim \N(z_{ij}-z_{ik}, 2\sigma^2),
\end{equation}
\small
\begin{equation}\label{lastone}
p_{jk} = \mathbb P(s_{ij}>s_{ik})=\mathbb P(s_{ij}-s_{ik}>0)=\mathbb P(d_{jk}>0)=\int_{0}^{\infty} h(d_{jk})d(d_{jk}),
\end{equation}
\normalsize
where $h$ denotes the density function. Then the probability that item $j$ ranks lower than item $k$, $p_{kj}=(1-p_{jk})$. The distribution of the rank $r_{ij}$ of item $j$ under the pair-wise contest approximation can be obtained by considering $r_{ij}$ as a Binomial-like random variable, i.e. equal to the number of successes in $(D_i-1)$ Bernoulli trial. Here by success we mean the case where another item beats item $j$ (ranked higher than $j$). Item $j$ is ranked in the $r_{ij}$ place, only if another $(r_{ij}-1)$ items beat it while it beats the rest $(D_i - r_{ij})$ items. The probability of success can be computed as above. For example, the probability that item $j$ is beaten by another item $k$, is $p_{kj}$, where $p_{kj}=(1-p_{jk})$ and $p_{jk}$ can be found in equation (\ref{lastone}). Accordingly, the probability that item $j$ ranked in position $r$ can be computed. In this way, $P(r_{ij}=r)$ can be calculated for each $r$ and then distribution of $r_{ij}$ is obtained.

Given that the discount function of $r_{ij}$ is as follow, the expectation of it can be calculated:
\begin{align}
\tilde{D}(r_{ij})= \frac{1}{\log_2(1+r_{ij})},\;\;\;\;\;\E(\tilde{D}(r_{ij})) = \sum_{r=1}^{D_i} D(r)\cdot P(r_{ij}=r).
\end{align}
Then we can replace the original discount function $D(r_{ij})$ with $\E(\tilde{D}(r_{ij}))$ and the adjusted NDCG of query $i$ becomes:
\begin{align}
\tilde{NDCG}(i) = G^{-1}_{max,i}\sum^{D_i}_{l=1}G_{il}\,\E(\tilde{D}(r_{il}))
\end{align}
Now this adjusted $\tilde{NDCG}$ can be directly taken as the objective function in the model training. A neural network model trained with this objective function is called Softrank.

\section{Results and Discussion}
\label{results}
\subsection{RQ1. Comparison to State-of-the-Art}
In this section we discuss the results of the comparison between the performance of each LTR model with its counterpart incorporating SIR, under a scale invariant test set (i.e., all features in the test set have consistent scale with those used for training). We used the average NDCG across all searches on the test set to measure the performance of the ranking algorithms. The corresponding validation and test NDCG can be found in Table \ref{scaleinvariantresult} ($2^{nd}$ and $3^{rd}$ columns). 

\begin{table*}[h]
    \centering
    \resizebox{0.7\linewidth}{!}{
    \setlength{\tabcolsep}{2mm}{
    \begin{tabular}{lllllll}
\hline
\textbf{Algorithm} & \textbf{Validation} & \textbf{Test}   & \textbf{Case 1}      & \textbf{Case 2}      & \textbf{Case 3}      & \textbf{Case 4}      \\ \hline
RankNet            & 0.684                    & 0.663                & 0.653                & 0.644                & 0.639                & 0.577                \\
RankNet (SIR)      & 0.683                    & 0.662                & 0.662***             & 0.662***             & 0.662***             & 0.662***             \\ \hline
LambdaRank         & 0.677                    & 0.656                & 0.647                & 0.639                & 0.634                & 0.573                \\
LambdaRank (SIR)   & 0.675                    & 0.655                & 0.655***             & 0.655***             & 0.655***             & 0.655***             \\ \hline
ListNet            & 0.526                    & 0.526                & 0.526                & 0.518                & 0.518                & 0.512                \\
ListNet (SIR)      & 0.510                    & 0.522                & 0.522                & 0.522***             & 0.522***             & 0.522***             \\ \hline
ListMLE            & 0.689                    & 0.668                & 0.654                & 0.647                & 0.639                & 0.570                \\
ListMLE (SIR)      & 0.686                    & 0.667                & 0.667***             & 0.667***             & 0.667***             & 0.667***             \\ \hline
SoftRank           & 0.682                    & 0.661                & 0.647                & 0.638                & 0.631                & 0.564                \\
SoftRank (SIR)     & 0.678                    & 0.660                & 0.660***             & 0.660***             & 0.660***             & 0.660***             \\\hline
                   & \multicolumn{1}{l}{}     & \multicolumn{1}{l}{} & \multicolumn{1}{l}{} & \multicolumn{1}{l}{} & \multicolumn{1}{l}{} & \multicolumn{1}{l}{}
\end{tabular}}}
    \caption{\small{RQ1 results: Validation NDCG during model training and Test NDCG are reported (for the scale invariant scenario) for RankNet, LambdaRank, ListNet, ListMLE, SoftRank and their ranking preserving counterparts (RankNet\textsubscript{SIR}, LambdaRank\textsubscript{SIR}, ListNet\textsubscript{SIR}, ListMLE\textsubscript{SIR}, and   SoftRank\textsubscript{SIR}).
RQ2 results: Test NDCG is reported for the four cases (Case 1-4) where we change the features' scale in the test set. Symbol *** on the table indicates the NDCG of SIR model is significant higher than its counterpart with p-value<0.001.}\vspace{-25pt}}
\label{scaleinvariantresult}
\end{table*}

As it can be seen from the $3^{rd}$ column of Table \ref{scaleinvariantresult}, all five LTR methods achieved comparable test set NDCG to their SIR counterpart (up to the third decimal point). 
In particular, ListMLE achieved the highest test set NDCG value, 0.668, followed by RankNet, SoftRank, and LambdaRank with 0.663 and 0.661 and 0.656 respectively. ListNet obtained the lowest test NDCG, 0.526. SIR models achieved comparable performance with a drop in NDCG at the third digit (-0.001 for RankNet\textsubscript{SIR}, LambdaRank\textsubscript{SIR}, ListMLE\textsubscript{SIR} and SoftRank\textsubscript{SIR}, and -0.004 for ListNet\textsubscript{SIR}). The validation NDCG during training has been also reported for completeness. 
The poor performance of ListNet is due to its loss formulation which gives relatively high weight to the negative samples \cite{listnet}. However, due to the large class imbalance in our data set (1 to 24), ListNet is not able to correctly learn the positive samples. Regarding RankNet and ListMLE, as mentioned in Section \ref{benchmarks}, to avoid noise, we only considered loss between items in different relevance levels and neglected ties. It could be shown that NDCG would decrease dramatically if taking loss between ties into consideration. Furthermore, owing to the large training set we have, NDCG for SoftRank would not increase significantly when including more than 8 negative samples, while the training time would increase exponentially in the number of used negative samples.

\vspace{10pt}
\noindent\fbox{%
    \parbox{\linewidth}{%
        \textbf{Answer to RQ1:} in the scenario where the training and test set features have the same scale, although the introduction of a scale invariant ranking function is reducing the expressivity of the model (i.e., limiting the non-linear interaction only among scale invariant features), it does not impact its prediction performance.
    }%
}

\subsection{RQ2. Comparison when scale in test set is perturbed}
In this section, we investigate how the methods perform when the test set features' scale is perturbed. Here we simulated four scenarios that could happen in a real-world e-commerce production environment, and report the test NDCG of the five LTR methods and their SIR counterpart (Table \ref{scaleinvariantresult}, $4^{th}$ to $7^{th}$ column). 
In our training set the price and discount features are recorded as the nightly values in a unified currency (i.e., USD). However, in the real-world e-commerce environment, for some point of sales, the price and discount are displayed as the full stay (i.e., the product of the nightly price/discount and the number of nights). In other words, the scale of the price and discount in the prediction phase is not always consistent with the one used for training. This situation is commonly seen due to differences in marketing strategies for different countries, and our goal is for the ranking system to perform consistently under this scale heterogeneity. In \textit{Case 1} we simulated this business scenario by multiplying the nightly price and discount on the test set by the number of nights. The results ($4^{th}$ column of Table \ref{scaleinvariantresult}) show that all SIR solutions maintain the same NDCG achieved in RQ1, while the original methods have a significant drop in NDCG. In particular, RankNet has a drop of -0.01, LambdaRank -0.009, ListMLE -0.014 and SoftRank -0.014. ListNet is the only method remaining the original NDCG, though as we have seen in previous results, this is the only model which struggled to learn the task at hand, and showed an NDCG which is more than 10 decimal points worse than all the other models. As already mentioned, while the price and discount currency is unified for training and analysis purposes, the one displayed on the production environment usually depends on the user choice. To simulate this (\textit{Case 2}, $5^{th}$ column of Table \ref{scaleinvariantresult}), we multiply  hotel prices and discounts in the test set by the local currency recorded when the search has been performed.
As before, SIR held the same NDCG registered in RQ1, while the methods not addressing the scale invariance had a significant drop in NDCG: RankNet has a drop of -0.019 , LambdaRank -0.017, ListNet -0.008, ListMLE -0.021 and SoftRank -0.023. In \textit{Case 3} ($6^{th}$ column of Table \ref{scaleinvariantresult}), we combined the changes over the first two scenarios, to get an even stronger shift in scale: the price and discount are passed to the model at prediction time as the full price/discount of the stay in the local currency. Our model was once again able to preserve the ranking with same test NDCG as the original case, while all the other methods had relatively lower NDCG (from -0.008 to -0.030). In detail, RankNet has a drop of -0.024, LambdaRank -0.022, ListNet -0.008, ListMLE -0.029 and SoftRank -0.030. 
In the last case ($7^{th}$ column of Table \ref{scaleinvariantresult}), we transformed the test set hotel price and discount to Korean Won by multiplying the original values (expressed in dollars) with the Korean Won daily exchange rate (i.e., 1 USD = \char`\~1200 KWR), to simulate the business scenario that all bookings on the test set were done from the Korean point of sale. As it can be seen also in this last case, all the models adopting SIR have achieved an identical NDCG to the one reported in RQ1, while the models not addressing the problem had a substantial drop in NDCG (up to -0.098). 
In particular, RankNet has a drop of -0.086 , LambdaRank -0.083, ListNet -0.014, ListMLE -0.098 and SoftRank -0.097.
As it can be seen in Table~\ref{scaleinvariantresult}, the NDCG achieved by the SIR models in the four simulated cases, is significantly higher (p-value $<0.001$) than their original counterparts in all but one case. The original formulation of ListNet achieved better NDCG in Case 1 (which is the one with the smallest perturbation compared to the test set), while still showing a consistent drop in NDCG in the remaining three cases. ListNet\textsubscript{SIR}, while being worse in Case 1, it still maintains the ranking preserving feature, and a consistent NDCG across all analysed scenarios.

As reported in \cite{cheng2016wide}, when considering big e-commerce with millions of daily users, even a 0.01 difference in the offline evaluation metric (e.g., NDCG) can lead to statistically significant changes in online metrics such as conversion rate \cite{mcfarland2012experiment} or user engagement \cite{drutsa2015practical}.

\vspace{10pt}
\noindent\fbox{%
    \parbox{\linewidth}{%
        \textbf{Answer to RQ2:} The proposed method is invariant to feature scale changes regardless of the magnitude. Furthermore, the original formulations are not ranking invariant, leading to a negative impact on the evaluation metric up to 14.7\% depending on the approach and the considered scenario.
    }%
}


\section{Threats to Validity}
Several factors can bias the validity of empirical studies. In this section we discuss the construct, conclusion and external validity threats that may affect our study.
To satisfy construct validity a study has ``to establish correct operational measures for the concepts being studied'' \cite{kitchenham1995case}. This means that the features and target variables should precisely measure the concepts they claim to measure. We mitigated such a threat by using a real-world dataset and excluded all the independent variables that are not known at inference time and therefore cannot be used for prediction purposes.
With regards to the conclusion validity, we carefully applied the statistical tests, verifying all the required assumptions. 
Moreover, to mitigate external validity threats we described the theorical framework behind the proposed method and used a large dataset spanning multiple years of data, cointaining 56 million booked searches. However we cannot claim that our results generalise beyond the subjects studied herein.

\section{Conclusion}
This paper has introduced and evaluated a scale invariant ranking function for Learing-To-Rank tasks. 
Results indicate that our proposal is rank preserving when the features' scale is perturbed at prediction time, maintaining the same NDCG performance. This feature is extremely important to make the Learning-To-Rank model robust in a production environment, when a sudden change in scale of a feature (e.g., due to A/B testing) could unexpectedly alter the ranking order. Our results also show that all original formulations are not rank preserving and their performance drops proportionally to the shift in the features' scale. Given our proposal is able to answer positively both our research questions, we would suggest practitioners in the LTR domain to address the scale variance problem when deploying ranking models to the production environment in order to avoid unwilling changes in ranking.

\bibliographystyle{unsrt}  
\bibliography{template}

\begin{thebibliography}{10}

\bibitem{adomavicius2011context}
Gediminas Adomavicius and Alexander Tuzhilin.
\newblock Context-aware recommender systems.
\newblock In {\em Recommender systems handbook}, pages 217--253. Springer,
  2011.

\bibitem{ling2014opening}
Liuyi Ling, Xiaolong Guo, and Chenchen Yang.
\newblock Opening the online marketplace: An examination of hotel pricing and
  travel agency on-line distribution of rooms.
\newblock {\em Tourism management}, 45:234--243, 2014.

\bibitem{xiang2015adapting}
Zheng Xiang, Dan Wang, Joseph~T O’Leary, and Daniel~R Fesenmaier.
\newblock Adapting to the internet: trends in travelers’ use of the web for
  trip planning.
\newblock {\em Journal of travel research}, 54(4):511--527, 2015.

\bibitem{petrozziello2017data}
Alessio Petrozziello and Ivan Jordanov.
\newblock Data analytics for online travelling recommendation system: a case
  study.
\newblock In {\em Proceedings of the IASTED International Conference Modelling,
  Identification and Control (MIC 2017), Innsbruck, Austria}, pages 106--112,
  2017.

\bibitem{guo2018novel}
Longhua Guo, Jianhua Li, Jie Wu, Wei Chang, and Jun Wu.
\newblock A novel airbnb matching scheme in shared economy using confidence and
  prediction uncertainty analysis.
\newblock {\em IEEE Access}, 6:10320--10331, 2018.

\bibitem{ai2019learning}
Qingyao Ai, Xuanhui Wang, Sebastian Bruch, Nadav Golbandi, Michael Bendersky,
  and Marc Najork.
\newblock Learning groupwise multivariate scoring functions using deep neural
  networks.
\newblock In {\em Proceedings of the 2019 ACM SIGIR International Conference on
  Theory of Information Retrieval}, pages 85--92, 2019.

\bibitem{ranknet}
Chris Burges, Tal Shaked, Erin Renshaw, Ari Lazier, Matt Deeds, Nicole
  Hamilton, and Greg Hullender.
\newblock Learning to rank using gradient descent.
\newblock In {\em Proceedings of the 22nd international conference on Machine
  learning}, pages 89--96, 2005.

\bibitem{lambdarank}
Christopher~JC Burges.
\newblock From ranknet to lambdarank to lambdamart: An overview.
\newblock {\em Learning}, 11(23-581):81, 2010.

\bibitem{lambdarank2}
Xuanhui Wang, Cheng Li, Nadav Golbandi, Michael Bendersky, and Marc Najork.
\newblock The lambdaloss framework for ranking metric optimization.
\newblock In {\em Proceedings of the 27th ACM International Conference on
  Information and Knowledge Management}, pages 1313--1322, 2018.

\bibitem{listnet}
Zhe Cao, Tao Qin, Tie-Yan Liu, Ming-Feng Tsai, and Hang Li.
\newblock Learning to rank: from pairwise approach to listwise approach.
\newblock In {\em Proceedings of the 24th international conference on Machine
  learning}, pages 129--136, 2007.

\bibitem{listmle}
Fen Xia, Tie-Yan Liu, Jue Wang, Wensheng Zhang, and Hang Li.
\newblock Listwise approach to learning to rank: theory and algorithm.
\newblock In {\em Proceedings of the 25th international conference on Machine
  learning}, pages 1192--1199, 2008.

\bibitem{softrank}
Michael Taylor, John Guiver, Stephen Robertson, and Tom Minka.
\newblock Softrank: optimizing non-smooth rank metrics.
\newblock In {\em Proceedings of the 2008 International Conference on Web
  Search and Data Mining}, pages 77--86, 2008.

\bibitem{sorokina2016amazon}
Daria Sorokina and Erick Cantu-Paz.
\newblock Amazon search: The joy of ranking products.
\newblock In {\em Proceedings of the 39th International ACM SIGIR conference on
  Research and Development in Information Retrieval}, pages 459--460, 2016.

\bibitem{karmaker2017application}
Shubhra~Kanti Karmaker~Santu, Parikshit Sondhi, and ChengXiang Zhai.
\newblock On application of learning to rank for e-commerce search.
\newblock In {\em Proceedings of the 40th International ACM SIGIR Conference on
  Research and Development in Information Retrieval}, pages 475--484, 2017.

\bibitem{grbovic2018real}
Mihajlo Grbovic and Haibin Cheng.
\newblock Real-time personalization using embeddings for search ranking at
  airbnb.
\newblock In {\em Proceedings of the 24th ACM SIGKDD International Conference
  on Knowledge Discovery \& Data Mining}, pages 311--320, 2018.

\bibitem{panteli2019recommendation}
Maria Panteli, Alessandro Piscopo, Adam Harland, Jonathan Tutcher, and
  Felix~Mercer Moss.
\newblock Recommendation systems for news articles at the bbc.
\newblock 2019.

\bibitem{dixon2011b}
Eleri Dixon, Emily Enos, and Scott Brodmerkle.
\newblock A/b testing of a webpage, July~5 2011.
\newblock US Patent 7,975,000.

\bibitem{cheng2016wide}
Heng-Tze Cheng, Levent Koc, Jeremiah Harmsen, Tal Shaked, Tushar Chandra,
  Hrishi Aradhye, Glen Anderson, Greg Corrado, Wei Chai, Mustafa Ispir, Rohan
  Anil, Zakaria Haque, Lichan Hong, Vihan Jain, Xiaobing Liu, and Hemal Shah.
\newblock Wide \& deep learning for recommender systems, 2016.

\bibitem{haldar2019applying}
Malay Haldar, Mustafa Abdool, Prashant Ramanathan, Tao Xu, Shulin Yang,
  Huizhong Duan, Qing Zhang, Nick Barrow-Williams, Bradley~C Turnbull,
  Brendan~M Collins, et~al.
\newblock Applying deep learning to airbnb search.
\newblock In {\em Proceedings of the 25th ACM SIGKDD International Conference
  on Knowledge Discovery \& Data Mining}, pages 1927--1935, 2019.

\bibitem{sculley2015hidden}
David Sculley, Gary Holt, Daniel Golovin, Eugene Davydov, Todd Phillips,
  Dietmar Ebner, Vinay Chaudhary, Michael Young, Jean-Francois Crespo, and Dan
  Dennison.
\newblock Hidden technical debt in machine learning systems.
\newblock In {\em Advances in neural information processing systems}, pages
  2503--2511, 2015.

\bibitem{iqbal2019production}
Murium Iqbal, Nishan Subedi, and Kamelia Aryafar.
\newblock Production ranking systems: A review.
\newblock {\em arXiv preprint arXiv:1907.12372}, 2019.

\bibitem{wang2013theoretical}
Yining Wang, Liwei Wang, Yuanzhi Li, Di~He, Wei Chen, and Tie-Yan Liu.
\newblock A theoretical analysis of ndcg ranking measures.
\newblock In {\em Proceedings of the 26th annual conference on learning theory
  (COLT 2013)}, volume~8, page~6, 2013.

\bibitem{liu2009learning}
Tie-Yan Liu et~al.
\newblock Learning to rank for information retrieval.
\newblock {\em Foundations and Trends{\textregistered} in Information
  Retrieval}, 3(3):225--331, 2009.

\bibitem{manning2008introduction}
Christopher~D Manning, Prabhakar Raghavan, and Hinrich Sch{\"u}tze.
\newblock {\em Introduction to information retrieval}.
\newblock Cambridge university press, 2008.

\bibitem{dalip2013exploiting}
Daniel~Hasan Dalip, Marcos~Andr{\'e} Gon{\c{c}}alves, Marco Cristo, and Pavel
  Calado.
\newblock Exploiting user feedback to learn to rank answers in q\&a forums: a
  case study with stack overflow.
\newblock In {\em Proceedings of the 36th international ACM SIGIR conference on
  Research and development in information retrieval}, pages 543--552, 2013.

\bibitem{mohan2011web}
Ananth Mohan, Zheng Chen, and Kilian Weinberger.
\newblock Web-search ranking with initialized gradient boosted regression
  trees.
\newblock In {\em Proceedings of the learning to rank challenge}, pages 77--89,
  2011.

\bibitem{xu2007adarank}
Jun Xu and Hang Li.
\newblock Adarank: a boosting algorithm for information retrieval.
\newblock In {\em Proceedings of the 30th annual international ACM SIGIR
  conference on Research and development in information retrieval}, pages
  391--398, 2007.

\bibitem{lan2014position}
Yanyan Lan, Yadong Zhu, Jiafeng Guo, Shuzi Niu, and Xueqi Cheng.
\newblock Position-aware listmle: A sequential learning process for ranking.
\newblock In {\em UAI}, pages 449--458, 2014.

\bibitem{pennock2000social}
David~M Pennock, Eric Horvitz, C~Lee Giles, et~al.
\newblock Social choice theory and recommender systems: Analysis of the
  axiomatic foundations of collaborative filtering.
\newblock In {\em AAAI/IAAI}, pages 729--734, 2000.

\bibitem{yan2013exploiting}
Surong Yan, Xiaolin Zheng, Deren Chen, and Yan Wang.
\newblock Exploiting two-faceted web of trust for enhanced-quality
  recommendations.
\newblock {\em Expert systems with applications}, 40(17):7080--7095, 2013.

\bibitem{yu2008collaborative}
Li~Yu and Xiaoping Yang.
\newblock Collaborative filtering recommendation based on preference order.
\newblock In {\em Research and Practical Issues of Enterprise Information
  Systems II}, pages 1567--1573. Springer, 2008.

\bibitem{lemire2005scale}
Daniel Lemire.
\newblock Scale and translation invariant collaborative filtering systems.
\newblock {\em Information Retrieval}, 8(1):129--150, 2005.

\bibitem{su2009survey}
Xiaoyuan Su and Taghi~M Khoshgoftaar.
\newblock A survey of collaborative filtering techniques.
\newblock {\em Advances in artificial intelligence}, 2009, 2009.

\bibitem{anand2010adaptive}
Deepa Anand and Kamal~K Bharadwaj.
\newblock Adaptive user similarity measures for recommender systems: a genetic
  programming approach.
\newblock In {\em 2010 3rd International Conference on Computer Science and
  Information Technology}, volume~8, pages 121--125. IEEE, 2010.

\bibitem{xie2015mathematical}
Hong Xie and John~CS Lui.
\newblock Mathematical modeling and analysis of product rating with partial
  information.
\newblock {\em ACM Transactions on Knowledge Discovery from Data (TKDD)},
  9(4):1--33, 2015.

\bibitem{resnick1994grouplens}
Paul Resnick, Neophytos Iacovou, Mitesh Suchak, Peter Bergstrom, and John
  Riedl.
\newblock Grouplens: an open architecture for collaborative filtering of
  netnews.
\newblock In {\em Proceedings of the 1994 ACM conference on Computer supported
  cooperative work}, pages 175--186, 1994.

\bibitem{sarwar2000application}
Badrul Sarwar, George Karypis, Joseph Konstan, and John Riedl.
\newblock Application of dimensionality reduction in recommender system-a case
  study.
\newblock Technical report, Minnesota Univ Minneapolis Dept of Computer
  Science, 2000.

\bibitem{traupman2004collaborative}
Jonathan Traupman and Robert Wilensky.
\newblock Collaborative quality filtering: Establishing consensus or recovering
  ground truth?
\newblock In {\em International Workshop on Knowledge Discovery on the Web},
  pages 73--86. Springer, 2004.

\bibitem{yi2019positively}
Mingyang Yi, Qi~Meng, Wei Chen, Zhi-ming Ma, and Tie-Yan Liu.
\newblock Positively scale-invariant flatness of relu neural networks.
\newblock {\em arXiv preprint arXiv:1903.02237}, 2019.

\bibitem{neyshabur2015path}
Behnam Neyshabur, Russ~R Salakhutdinov, and Nati Srebro.
\newblock Path-sgd: Path-normalized optimization in deep neural networks.
\newblock In {\em Advances in Neural Information Processing Systems}, pages
  2422--2430, 2015.

\bibitem{yuan2019scaling}
Qunyong Yuan and Nanfeng Xiao.
\newblock Scaling-based weight normalization for deep neural networks.
\newblock {\em IEEE Access}, 7:7286--7295, 2019.

\bibitem{salimans2016weight}
Tim Salimans and Durk~P Kingma.
\newblock Weight normalization: A simple reparameterization to accelerate
  training of deep neural networks.
\newblock In {\em Advances in neural information processing systems}, pages
  901--909, 2016.

\bibitem{ba2016layer}
Jimmy~Lei Ba, Jamie~Ryan Kiros, and Geoffrey~E Hinton.
\newblock Layer normalization.
\newblock {\em arXiv preprint arXiv:1607.06450}, 2016.

\bibitem{liao2016streaming}
Qianli Liao, Kenji Kawaguchi, and Tomaso Poggio.
\newblock Streaming normalization: Towards simpler and more
  biologically-plausible normalizations for online and recurrent learning.
\newblock {\em arXiv preprint arXiv:1610.06160}, 2016.

\bibitem{raschka2018model}
Sebastian Raschka.
\newblock Model evaluation, model selection, and algorithm selection in machine
  learning.
\newblock {\em arXiv preprint arXiv:1811.12808}, 2018.

\bibitem{bergstra2013hyperopt}
James Bergstra, Dan Yamins, and David~D Cox.
\newblock Hyperopt: A python library for optimizing the hyperparameters of
  machine learning algorithms.
\newblock In {\em Proceedings of the 12th Python in science conference},
  volume~13, page~20. Citeseer, 2013.

\bibitem{liu2011learning}
Tie-Yan Liu.
\newblock {\em Learning to rank for information retrieval}.
\newblock Springer Science \& Business Media, 2011.

\bibitem{mcfarland2012experiment}
Colin McFarland.
\newblock {\em Experiment!: Website conversion rate optimization with A/B and
  multivariate testing}.
\newblock New Riders, 2012.

\bibitem{drutsa2015practical}
Alexey Drutsa, Anna Ufliand, and Gleb Gusev.
\newblock Practical aspects of sensitivity in online experimentation with user
  engagement metrics.
\newblock In {\em Proceedings of the 24th ACM International on Conference on
  Information and Knowledge Management}, pages 763--772, 2015.

\bibitem{kitchenham1995case}
Barbara Kitchenham, Lesley Pickard, and Shari~Lawrence Pfleeger.
\newblock Case studies for method and tool evaluation.
\newblock {\em IEEE software}, 12(4):52--62, 1995.

\end{thebibliography}


\end{document}